\documentclass[showpacs,floatfix,superscriptaddress]{revtex4}
\usepackage{graphicx}
\usepackage{multirow}
\usepackage{epsfig}
\usepackage{bm}
\usepackage[utf8]{inputenc}
\usepackage{amssymb}
\usepackage{float}
\usepackage{amsmath}
\usepackage{dcolumn}
\usepackage{latexsym}
\usepackage{subfig}
\usepackage{amsthm}
\usepackage{array}
\usepackage{framed}
\usepackage{cancel}
\usepackage[colorlinks]{hyperref}
\usepackage[usenames,dvipsnames]{color}
\usepackage[dvipsnames]{xcolor}
\hypersetup{
     breaklinks=true,
    pdfstartview={FitH},    
    colorlinks=true,       
    linkcolor=blue,          
    citecolor=red,        
    filecolor=magenta,      
    urlcolor=blue,           
    anchorcolor=green,      
    linktocpage=true
}

\def\doi{http://doi.org}

\begin{document}

\title{Investigating QED Effects on the Thin Accretion Disk Properties Around Rotating Euler-Heisenberg Black Holes}

\author{Kourosh Nozari}
\email[]{knozari@umz.ac.ir (Corresponding Author)}
\author{Sara Saghafi}
\email[]{s.saghafi@umz.ac.ir}
\author{Fatemeh Aliyan}
\email[]{f.aliyan03@umz.ac.ir}
\affiliation{Department of Theoretical Physics, Faculty of Science, University of Mazandaran,\\
P. O. Box 47416-95447, Babolsar, Iran}

\begin{abstract}
The Einstein-Euler-Heisenberg (EEH) black hole model represents an extension of classical black hole solutions in general relativity by incorporating quantum electrodynamic (QED) corrections. These corrections are introduced through the inclusion of the Euler-Heisenberg Lagrangian, which accounts for the nonlinear effects of QED in the presence of strong electromagnetic fields. This study investigates the observational properties of a thin accretion disk surrounding the electrically charged rotating EEH black hole. By exploring the influence of the spin parameter and charge on key dynamical quantities—such as the energy, angular momentum, angular velocity, and the innermost stable circular orbit (ISCO) of a test particle—it becomes possible to analyze the radiative flux, temperature distribution, and differential luminosity of the thin accretion disk in the spacetime of the charged rotating EEH black hole. The results are compared to those of Kerr and Kerr-Newman black holes in General relativity, revealing that QED corrections are found to increase the ISCO radius. Specifically, for a fixed electric charge, an increasing spin parameter leads to a larger ISCO radius compared to the standard Kerr black holes, as a result of additional electromagnetic corrections introduced by the Euler-Heisenberg theory. Conversely, when the spin parameter is held constant, an increase in the electric charge reduces the ISCO radius. Additionally, thin accretion disks around charged EEH rotating black holes exhibit higher temperatures and greater efficiency when the spin parameter is fixed and the electric charge is increased. However, for a constant electric charge, increasing the spin parameter results in an accretion disk that is cooler and has lower radiant efficiency. These findings highlight the potential of accretion disk processes as valuable tools for probing Euler-Heisenberg theory through astrophysical observations.
\vspace{12 pt}
\\


\end{abstract}


\maketitle

\enlargethispage{\baselineskip}
\tableofcontents

\section{Inroduction}

Black holes are solutions to Einstein's field equations in General Relativity (GR) that describe regions of space-time undergoing intense gravitational collapse towards a classically one-way event horizon. Convincing evidence for the existence of black holes in astronomical observations has emerged from two major breakthroughs. The first significant discovery was made by the Laser Interferometer Gravitational-Wave Observatory (LIGO), which detected gravitational waves resulting from the coalescence of two black holes \cite{1}. This observation provided indirect but robust confirmation of the existence of black holes, consistent with GR predictions. Another groundbreaking achievement came from the Event Horizon Telescope (EHT) collaboration, which captured the first near-horizon image of the supermassive black hole at the center of the Messier 87 (M87$^{*}$) elliptical galaxy. This black hole has an estimated mass of $( M = (6.5 \pm 0.7) \times 10^{9} M_{\odot})$. The image revealed a compact, asymmetric ring-like structure, with a bright ring of radiation surrounding a circular dark region that corresponds to the black hole's shadow \cite{2,3,4,5,6,7}. This observation provided direct visual evidence for the existence of black holes in the Universe. Moreover, the EHT measured the linear polarization of the shadow of M87$^{*}$, offering critical insights into the magnetic field structure responsible for synchrotron emission. The results indicated that the black hole is surrounded by a magnetically arrested accretion disk, with an estimated mass accretion rate of $((3-20) \times 10^{-4} M_{\odot} \, \text{yr}^{-1})$\cite{8,9}. These magnetic fields play a pivotal role in launching the energetic jets observed at the core of M87$^{*}$. Recently, the EHT also captured the first horizon-scale radio observations of Sagittarius A$^{*}$ (Sgr A$^{*}$), the supermassive black hole at the center of the Milky Way. The observed ring size of Sgr A$^{*}$ agrees with the shadow critical curve predicted by GR within 10\% \cite{eht-1,eht-2,eht-3,eht-4,eht-5,eht-6}. This finding further reinforces the predictions of GR and enhances our understanding of black hole physics.\\

Astrophysical bodies such as black holes typically grow in mass through accretion. When interstellar material gathers around a compact object, it usually organizes into an accretion disk—a rotating, disk-shaped structure of gas gradually spiraling toward the central mass. As the gas moves deeper into the gravitational well of the compact object, it loses gravitational energy, which is predominantly transformed into thermal energy. A portion of this heat is radiated away, particularly from the inner regions of the accretion disk, allowing it to cool over time. This emitted radiation spans various wavelengths, including radio, optical, and X-ray, and can be detected using corresponding telescopes. Analyzing the electromagnetic spectrum of this radiation provides critical insights into the dynamics of accretion disks. The properties of the observed radiation are intimately linked to the geodesic trajectories of the gas particles, which are themselves shaped by the geometry and nature of the central massive object. Therefore, analyzing the emission spectra of accretion disks provides crucial insights into the astrophysical characteristics of these compact bodies. Recent advancements in observational astronomy have substantially increased the precision of accretion disk analysis. These improvements now allow for detailed measurements of critical observables, such as temperature and luminosity. As a result, they offer indispensable empirical support for testing black hole models that extend beyond the scope of General Relativity.\\

The classical model for geometrically thin accretion disks, originally proposed by Shakura and Sunyaev in 1973 \cite{Shakura}, was first developed within the Newtonian framework and subsequently extended to a relativistic context by Novikov and Thorne \cite{Novikov}. This model assumes a steady-state configuration where the mass accretion rate remains constant across the disk, regardless of radial distance. It also presumes that the accreting matter follows Keplerian orbits, which implies that the central compact object lacks a strong magnetic field. Furthermore, the radiation emitted by the accretion disk is modeled as blackbody radiation, resulting from the thermodynamic equilibrium of the disk material. The spatial distribution of energy flux over the disk surface, along with the radiative efficiency—defined as the portion of the infalling matter's rest mass energy converted into radiation—were thoroughly investigated in \cite{Page} and \cite{Thorne}.
. Thin accretion disk models have been widely investigated in the context of modified gravity theories, including $f(R)$ gravity \cite{FR1}--\cite{FR3}, scalar-tensor-vector gravity \cite{SVT}, Einstein-Maxwell-dilaton theory \cite{EMd1}--\cite{EMd2}, Einstein-scalar-Gauss-Bonnet gravity \cite{EdGB1}--\cite{EdGB2}, Chern-Simons gravity \cite{Chern}, Einstein-Gauss-Bonnet gravity \cite{LIU,Heydari-Fard:2021ljh} and Horava-Lifshitz gravity \cite{Horava}. Similarly, such disks have been studied in higher-dimensional gravity frameworks, such as Kaluza-Klein and brane-world models, as detailed in \cite{Kaluza}--\cite{brane2}. Research has also extended the Novikov-Thorne model to analyze disks around wormholes, neutron stars, boson stars, fermion stars, and naked singularities, as explored in \cite{WH1}--\cite{nk3}. The analysis of the accretion disk surrounding the Euler-Heisenberg Anti-de Sitter black hole has been conducted in \cite{Abbas:2023nra}. In this paper, we seek to extend the analysis to encompass the case of rotating black hole solutions within the framework of EEH gravity, as astrophysical black holes are anticipated to be rapidly rotating due to the effects of accretion. \\

One of the most challenging aspects of GR is the presence of singularities, which is manifested at the beginning of the Universe and at the center of black holes. Maxwell's equations also exhibit singularities, leading to divergence issues within classical electrodynamics. To address these challenges, Euler and Heisenberg introduced a novel approach to describing the electromagnetic field, grounded in Dirac's positron theory. This approach incorporates one-loop corrections in QED, providing a theoretical framework to explain vacuum polarization effects in QED~\cite{26}. Nonlinear electrodynamic models have also been utilized to describe early Universe inflation~\cite{27}. Yajima et al.~derived the Euler-Heisenberg black hole solution by incorporating the one-loop effective Lagrangian density into the Einstein field equations~\cite{28} and the inclusion of electric charge was addressed in Ref. \cite{Ruffini:2013hia}. The charged static configuration was subsequently extended to an axisymmetric case in Ref. \cite{Breton:2019arv} using the Newman-Janis algorithm, and the rotating structure of the black hole was further analyzed in Ref. \cite{Breton:2022}. The motion of time-like particles in the EEH framework was explored in Ref. \cite{Amaro:2023ull}, while numerous studies on the non-rotating case of EEH black holes can be found in the literature \cite{Zeng:2022pvb,Breton:2021mju,Luo:2022gdz,Dai:2022mko,Feng:2022otu,
Maceda:2018zim,Maceda:2020rpv,Rehman:2023hro,Mushtaq:2024cap}. In \cite{Abbas:2023nra} the authors investigate the effects of EEH theory in AdS backgrounds and provides a complementary view of disk structur. Another
relevant work is \cite{Abbas:2023rzk} , which examines ISCO, radiative
flux, and efficiency in modified gravity. Furthermore, Abbas and collaborators offers key insights into particle dynamics and disk physics
in the presence of nonlinear electromagnetic fields \cite{Ditta:2023vou}, and also in a recent work \cite{Rehman:2025rww}, they broadens the EEH framework to include cosmological fields. Additionally, Kruglov determined the effective geometry and shadow size of non-rotating magnetic black holes by incorporating nonlinear electrodynamics corrections~\cite{Kruglov-2}.\\

Despite significant advancements in the study of black hole physics within the framework of nonlinear electrodynamics, there remains a significant gap in understanding of their effects on thin accretion disks of rotating black holes in EEH theory. This study will investigate the observational characteristics of a thin accretion disk surrounding the electrically charged rotating EEH black hole derived in \cite{Breton:2019arv}. In particular, we will explore how QED effects influence the properties and observables of thin accretion disks. This research aims to provide a foundation for future studies focused on evaluating alternatives to General Relativity through accretion disk observations. The structure of the paper is as follows: Section II revisits the EEH theory, outlining its formulation using dual Plebański variables and reviewing the electrically charged rotating black hole solution within this theoretical framework. Section III explores the geodesic motion of massive particles confined to the equatorial plane, with particular focus on the influence of the rotation parameter $a$ and electric charge $Q$. In Section IV, the radiative characteristics of a thin accretion disk surrounding the electrically charged rotating EEH black hole are investigated, emphasizing the numerical analysis of how parameters $a$ and $Q$ affect these properties. Finally, Section V presents the conclusions and a discussion of the results.

\section{ Einstein-Euler-Heisenberg theory and its electrically charged rotating black hole solution }\label{se1}

The action describing Einstein gravity minimally coupled to the Euler–Heisenberg (EH) theory is given by \cite{HieEu,GibsonW}
\begin{equation}\label{eq1}
  \mathcal{S}=\frac{1}{4\pi G}\int_{M^{4}}d^{4}x\sqrt{-g}\Big(\frac{1}{4}R+[-X+\frac{A}{2}X^2+\frac{B}{2}Y^2]\Big)\,.
\end{equation}
In this equation, $R$ represents the Ricci scalar curvature, $G$ is the Newton's constant which we set to unity, and $X$ and $Y$ are the only two independent relativistic invariant and pseudo-invariant variables from the Maxwell field in four dimensions.
\begin{equation}\label{eq2}
X=\frac{1}{4}\mathcal{F}_{\mu\upsilon}\mathcal{F}^{\mu\upsilon}\,,
Y=\frac{1}{4}\mathcal{F}_{\mu\upsilon}{^{*}\mathcal{F}^{\mu\upsilon}},
\end{equation}
$A = \frac{8\alpha^{2}}{45m^4}$, $B = \frac{7\alpha^{2}}{180m^4 }= \frac{7}{4} A$, where $m$ represents electron mass and $\alpha$ is the fine structure constant. $*\mathcal{F}_{\mu\upsilon}$ is the dual of the Faraday tensor $\mathcal{F}_{\mu\upsilon} = A_{\mu;\upsilon}-A_{\upsilon;\mu}$, which is defined as usual $^{*}\mathcal{F}_{\mu\upsilon} = \frac{1}{2} \epsilon_{\mu\upsilon\sigma\rho} \mathcal{F}^{\sigma\rho}$, where $\epsilon_{\mu\upsilon\sigma\rho}$ is the totally antisymmetric tensor with $\epsilon_{\mu\upsilon\sigma\rho}\epsilon^{\mu\upsilon\sigma\rho} = -4!$.

The Legendre dual model of non-linear electrodynamics, proposed by Pleba'nski in \cite{PL}, simplifies the equations of motion by introducing the tensor $\mathcal{P}_{\mu\upsilon} = B_{\mu;\upsilon}-B_{\upsilon;\mu}$ defined as
\begin{equation}\label{eq3}
 d\mathcal{L}(X,Y)=-\frac{1}{2}\mathcal{P}^{\mu\upsilon}d\mathcal{F}_{\mu\upsilon}\,.
\end{equation}
The Lagrangian density for Euler-Heisenberg non-linear electrodynamics is $\mathcal{L}(X,Y)=-X+\frac{A}{2}X^2+\frac{B}{2}Y^2$. It's worth noting that $\mathcal{P}_{\mu\upsilon}$ coincide with $\mathcal{F}_{\mu\upsilon}$ for the linear Maxwell theory. In general it is:
\begin{equation}\label{eq4}
 \mathcal{P}_{\mu\upsilon}=-(\mathcal{L}_{X}\mathcal{F}_{\mu\upsilon}+Y^{} {^{*}\mathcal{F}_{\mu\upsilon}})\,,
\end{equation}
subscripts on $\mathcal{L}$ denote differentiation. In this situation, it reads
\begin{equation}\label{eq5}
\mathcal{P}_{\mu\upsilon}=-[(-1+AX)\mathcal{F}_{\mu\upsilon}+BY {^{*}\mathcal{F}_{\mu\upsilon}}]\,.
\end{equation}
Equation \eqref{eq4} represents the constitutive relations of the Euler-Heisenberg non-linear electrodynamics theory. The components of $\mathcal{P}_{\mu\upsilon}$ are the electric induction $D$ and magnetic field $H$. We use $s$ and $t$ to represent independent and pseudo-invariant Plebański variables $\mathcal{P}_{\mu\upsilon}$, which are defined as follows
\begin{equation}\label{eq6}
\begin{split}
& s=-\frac{1}{4}\mathcal{P}_{\mu\upsilon}\mathcal{P}^{\mu\upsilon}\,,\\
& t=-\frac{1}{4}\mathcal{P}_{\mu\upsilon} {^{*}\mathcal{P}^{\mu\upsilon}}\,.\\
\end{split}
\end{equation}
Where $ ^{*}\mathcal{P}_{\mu\upsilon}=\frac{1}{2}\epsilon_{\mu\upsilon\sigma\rho}\mathcal{P}^{\sigma\rho}$.

The covariant Hamiltonian $H(s, t)$ can be described as
\begin{equation}\label{eq7}
H(s,t)=-\frac{1}{2}\mathcal{P}^{\mu\upsilon}\mathcal{F}_{\mu\upsilon}-\mathcal{L}\,.
\end{equation}

In the Euler-Heisenberg theory, the Hamiltonian (up to first order in constants $A$ and $B$) reads
\begin{equation}\label{eq8}
H(s,t)=s-\frac{A}{2}s^2-\frac{B}{2}t^2\,.
\end{equation}

The coupled system's equations of motions are now \cite{SAlz}
\begin{equation}\label{eq9}
\begin{split}
& \nabla_{\mu}\mathcal{P}^{\mu\upsilon}=0\,,\\
& R_{\mu\upsilon}-\frac{1}{2}Rg_{\mu\upsilon}=8\pi T_{\mu\upsilon}\,,\\
\end{split}
\end{equation}
with
\begin{equation}\label{eq10}
T_{\mu\upsilon}=\frac{1}{4 \pi}[H_{s}\mathcal{P}_{\mu}^{\beta}\mathcal{P}_{\upsilon\beta}+g_{\mu\upsilon}(2sH_{s}+tH_{t}-H)]\,.
\end{equation}
The energy-momentum tensor for the Euler-Heisenberg non-linear electromagnetic field is
\begin{eqnarray}\label{tmunu}
T_{\mu\upsilon}& =&\frac{1}{4 \pi}[(1-As)\mathcal{P}_{\mu}^{\beta}\mathcal{P}_{\upsilon\beta}+g_{\mu\upsilon}(s-\frac{3}{2}As^{2}-\frac{B}{2}t^2)] \nonumber \\
&& =\frac{1}{4 \pi}[(-1-2AX)\mathcal{F}_{\mu}^{\beta}\mathcal{F}_{\upsilon\beta}-BY(\mathcal{F}_{\mu}^{\beta} {^{*}\mathcal{F}_{\upsilon\beta}}
+^{*}\mathcal{F}_{\mu}^{\beta}\mathcal{F}_{\upsilon\beta})+g_{\mu\upsilon}(s-\frac{3}{2}As^{2}-\frac{B}{2}t^2)]\,.
\end{eqnarray}
When $A=B=0$, Eq. \eqref{tmunu} yields the usual linear Maxwell energy-momentum tensor.\\
To recover the original variables, we employ constitutive or material equations that connect $\mathcal{F}_{\mu\upsilon}$ with $\mathcal{P}_{\mu\upsilon}$, which read
\begin{equation}\label{eq10}
\mathcal{F}_{\mu\upsilon}=[H_{s}+H_{t}]\mathcal{P}_{\mu\upsilon}=(1-As-Bt)\mathcal{P}_{\mu\upsilon}\,.
\end{equation}
For the electrically charged solution, one use the Pleba'nski dual variables of the non-linear electromagnetic field
\begin{equation}\label{eq11}
\mathcal{P}_{\mu\upsilon}=\frac{Q}{r^2}\Big(\delta_{\mu}^{0} \delta_{\upsilon}^{1}-\delta_{\mu}^{1} \delta_{\upsilon}^{0}\Big)\,.
\end{equation}
By considering the following static and spherically symmetric black hole metric
\begin{equation}\label{eq12}
ds^{2}=-\Big(1-\frac{2m(r)}{r}\Big)dt^{2}+\Big(1-\frac{2m(r)}{r}\Big)^{-1}dr^{2}+r^2\Big(d\theta^{2}+sin^{2}\theta d\phi^{2}\Big)\,,
\end{equation}
then by applying Eq. \eqref{eq11} and from the component $(0,0)$ of the Einstein’s equations one obtains
\begin{equation}\label{eq12}
m(r)=M-\frac{Q^2}{2r}+A \frac{Q^4}{4r^5}\,,
\end{equation}
in which $A$ is the EH parameter. Next, Lorentz covariant solutions for spinning systems are obtained, representing a relativistic variation of the Newman-Janis technique. The solution of electrically charged EEH rotating black hole are obtained as follows \cite{Breton:2019arv}
\begin{equation}\label{metric}
d s^{2}=g_{t t} d t^{2}+g_{r r} d r^{2}+g_{\theta \theta} d \theta^{2}+g_{\phi \phi} d \phi^{2}+2 g_{t \phi} d t d \phi,
\end{equation}
with the metric components are given by
\begin{align}\label{mc}
g_{t t}&= -(1-\frac{2Mr-Q^{2}+AQ^{4}/20r^{4}}{\rho^{2}})\quad   g_{t\phi}=-\frac{4\textmd{a}\Big(Mr-Q/2+(AQ^{4})/(40r^{4})\Big)\sin\theta^{2}}{\rho^{2}} \nonumber\\
g_{r r} &=\frac{\rho^{2}}{\Delta}\quad   g_{\theta \theta} =\rho^{2}\quad   g_{\phi\phi}=\frac{\Sigma\sin^{2}\theta}{\rho^{2}},
\end{align}
where
\begin{equation}\label{eq20}
  \rho^{2}=r^{2}+\textmd{a}^{2}\cos^{2}\theta
\end{equation}
\begin{equation}\label{eq21}
  \Delta=r^{2}+\textmd{a}^{2}-2m(r)r
\end{equation}
\begin{equation}\label{eq22}
  \Sigma=(r^{2}+\textmd{a}^{2})^{2}-\textmd{a}^{2}\Delta\sin^{2}\theta.
\end{equation}
Setting $\textmd{a} =0$ recovers the static screened Reissner-Nordstrom solution and also setting the EH parameter $A=0$, recovers the Kerr-Newman solution. To get physical insight into the energy-mass function, consider allowing it to vary over spacetime. In this paradigm, the solution behaves asymptotically as the Kerr-Newman one.

\section{ Geodesic equation and circular orbits of massive particles in EEH Black hole spacetime}\label{se3}

The motion of particles or photons in the spacetime of an electrically charged, rotating EEH black hole—analogous to the Kerr solution—is characterized by three conserved quantities \cite{geo}: the energy $E$, the angular momentum along the axis of rotation $L$, and the Carter constant $\kappa$. It is convenient to express the angular momentum and Carter constant in terms of energy-scaled variables, defined as $\lambda\equiv\frac{L}{E}$ and $\eta\equiv\frac{\kappa}{E}$. The geodesic equations that describe the path of a test particle or photon in the spacetime given by Eq.~(\ref{mc}) can be written as follows

\begin{align}
\frac{\rho^{2}}{E} p^{r}&= \pm_{r} \sqrt{\mathcal{R}(r)}, \label{pr}\\
\frac{\rho^{2}}{E} p^{\theta} &= \pm_{\theta} \sqrt{\Theta(\theta)},  \label{p8}\\
\frac{\rho^{2}}{E} p^{\phi}&=\frac{\textmd{a}}{\Delta}\left(r^{2}+\textmd{a}^{2}-\textmd{a} \lambda\right)+\frac{\lambda}{\sin ^{2} \theta}-\textmd{a} , \label{pp}\\
\frac{\rho^{2}}{E} p^{t}&=\frac{r^{2}+\textmd{a}^{2}}{\Delta}\left(r^{2}+\textmd{a}^{2}-\textmd{a} \lambda\right)+\textmd{a}\left(\lambda-\textmd{a} \sin ^{2} \theta\right) ,\label{pt}
\end{align}
Here, $p^{\nu}(\nu=r,\theta,\phi,t)$ denote the components of the four-momentum, while $\pm_{r}$ and $\pm_{\theta}$ indicate the respective signs of $p^r$ and $p^{\theta}$. The functions $\mathcal{R}(r)$ and $\Theta(\theta)$ correspond to the radial and angular potentials, respectively, and are defined as follows

\begin{align}
\mathcal{R}(r)&=\left(r^{2}+\textmd{a}^{2}-\textmd{a} \lambda\right)^{2}-\Delta\left[\mu^2 r^2+\eta+(\lambda-\textmd{a})^{2}\right], \\
\Theta(\theta)&=\eta+\textmd{a}^{2} \cos ^{2} \theta-\lambda^{2} \cot ^{2} \theta .
\end{align}
For photons, $\mu = 0$, while for massive particles, $\mu = 1$, indicating that the conserved quantities $E$, $\lambda$, and $\eta$ are normalized with respect to the particle’s mass $\mu$. In this study, we concentrate on the conserved quantities associated with time-like geodesics, normalized per unit mass.

We assume that all particles within the thin accretion disk are massive and follow circular orbits confined to the equatorial plane, moving along time-like geodesics. In order to analyze the observational properties of such disks, it is essential to first understand certain fundamental parameters describing the motion of these particles. Although our focus is on circular motion in the equatorial plane—where $\kappa = 0$ and $\theta = \pi/2$—we begin by presenting the general expressions for the relevant quantities. These general forms are more transparent and broadly applicable than their specific counterparts.

In general, assuming that the metric coefficients do not depend on the $t$ and $\phi$ coordinates, the Euler-Lagrange equations yield two conserved quantities: the energy per unit rest mass, $E$, and the angular momentum per unit rest mass, $L$,
\begin{equation}\label{eq24}
g_{tt}\dot{t}+g_{t\phi}\dot{\phi}=-E\,,
\end{equation}
\begin{equation}\label{eq25}
g_{t\phi}\dot{t}+g_{\phi\phi}\dot{\phi}=L\,.
\end{equation}
Using the equations above and the assumption $2{\cal L}=-1$, we can find
\begin{equation}
g_{rr} \dot{r}^2+g_{\theta\theta}\dot{\theta}^2=V_{\rm eff}(r,\theta),
\label{9}
\end{equation}
the effective potential is provided by
\begin{equation}\label{eq29}
  V_{eff}(r,\theta)=-1+\frac{E^{2}g_{\phi\phi}+2ELg_{t\phi}+L^{2}g_{tt}}{g_{t\phi}^{2}-g_{tt}g_{t\phi}}.
\end{equation}

\begin{figure}[htb]
\centering
\includegraphics[width=0.7\textwidth]{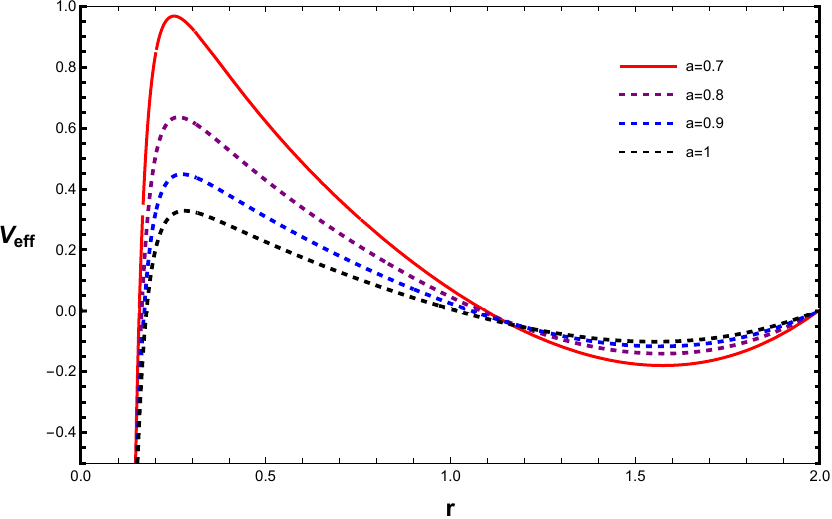}
\caption{\label{fig1}\small{\emph{The effective potential as function of the radial coordinate $r$ for the electrically charged rotating EEH black hole with the EH parameter $A=0.99,Q=0.1$ and different values of the spin parameter.}}}
\end{figure}

Effective potential analysis plays a major role in investigation of the circular orbits. In Fig .~(\ref{fig1}) we plot the effective potential $V_{eff}$ for the particle in the electrically charged rotating EEH black hole background as a function of radial coordinate $r$. Right panel illustrates the behavior of the
effective potential  for the electrically charged EEH rotating black hole in comparison with the Kerr and Kerr-newman black hole for different values of $Q$ by fixing the spin parameter. The left panel shows the curve of the $V_{eff}$ verses $r$ for different values of spin parameter by fixing $Q$. The position of circular orbits can be established by local extremum of the effective potential i.e. $ V_{eff}(r,\theta)=0$ , $\frac{d V_{eff}(r,\theta)}{dr}=\frac{d V_{eff}(r,\theta)}{d\theta}=0$. In the Novikov-Thorne model, the inner boundary of thin accretion disk is determined by $r_{\rm isco}$, as equatorial circular orbits are unstable when $r<r_{\rm isco}$. Consider the particle in the equatorial plane ($\theta=\frac{\pi}{2}$), moving along circular orbit whit $\dot{r}=\dot{\theta}=0$, and that $\ddot{r}=\ddot{\theta}=0$ if the particle is located in the innermost stable circular orbit known as the ISCO radius. Then we have
\begin{equation}\label{30}
 \frac{d^{2}V_{eff}}{dr^{2}}=\frac{1}{g_{t\phi}^{2}-g_{tt}g_{\phi\phi}}[E^{2}g_{\phi\phi,rr}+2ELg_{t\phi,rr}+L^{2}g_{tt,rr}-(g_{t\phi}^{2}-g_{tt}g_{\phi\phi})_{,rr}]=0\,.
\end{equation}
This equation represents the second radial derivative of the effective potential $( V_{eff} )$ for a test particle moving in a stationary, axisymmetric spacetime. The condition 
$
\frac{d^2 V_{eff}}{dr^2} = 0$, typically appears in the study of circular orbits, particularly when analyzing their stability. But, this equation cannot be solved analytically. Therefore we numerically calculated the ISCO radius for different values of parameters $\textmd{a}$ and $Q$ in Table \ref{t1}. In the case of $a=Q=0$, the equation above reduces to the Schwarzschild black hole with $r_{\rm isco}=6M$. For the Kerr black hole by setting $\textmd{a}=0.3$ we have $r_{\rm isco}=4.98M$.\\

\begin{table*}
 \caption{Numerical values of $r_{\rm{isco}}$, maximum energy flux ($F_{\rm{max}}$), maximum temperature ($T_{\rm{max}}$), and efficiency ($\epsilon$) of an accretion disk of the electrically charged rotating EEH black hole (we set $M=1$). As it is shown, $A=0$ and $Q=0$ corresponds to Kerr black hole, $A=0$ corresponds to Kerr-Newman black hole.}
 \label{t1}
 \begin{tabular}{ccccccc}
  \hline
  $A$ &$\textmd{a}$ &\hspace{5mm} $Q$ &\hspace{5mm} $r_{\rm{isco}}$ &\hspace{5mm} $F_{\rm{max}}$ ($\times10^{-6}$) & \hspace{5mm}$T_{\rm{max}}$ & \hspace{5mm}$\epsilon \%$ \\
  \hline
 & \hspace{5mm} & \hspace{5mm} 0.1 &\hspace{5mm} 7.70 & 1.952 & 0.0371 &\hspace{5mm} 4.414 \\
 & \hspace{5mm} & \hspace{5mm} 0.2 &\hspace{5mm} 7.66 & 1.993 & 0.0376 &\hspace{5mm} 4.435 \\
 $0.99$ & 0.3 &\hspace{5mm}  0.3 &\hspace{5mm} 7.60 & 2.066 & 0.0379 &\hspace{5mm} 4.471 \\
 & \hspace{5mm} &\hspace{5mm}  0.4 &\hspace{5mm} 7.50 & 2.205 & 0.038 &\hspace{5mm} 4.523 \\
 & \hspace{5mm} &\hspace{5mm}  0.5 &\hspace{5mm} 7.37 & 2.352 & 0.039 &\hspace{5mm} 4.593 \\
 $0$ & 0.3 &\hspace{5mm}  0 &\hspace{5mm} 4.98 & 1.89 & 0.0365 &\hspace{5mm} 6.94 \\
 $0$  & 0.3 &\hspace{5mm}  0.3 &\hspace{5mm} 2.58 & 1.93 & 0.0374 &\hspace{5mm} 7.1 \\
  \hline
  & \hspace{5mm}  0.3 &\hspace{5mm}  &\hspace{5mm} 7.60 & 2.067 & 0.038 &\hspace{5mm} 4.471 \\
  & \hspace{5mm}  0.5 &\hspace{5mm}  &\hspace{5mm} 8.59 & 1.144 & 0.032 &\hspace{5mm} 3.941 \\
  $0.99$ & \hspace{5mm}  0.8 &\hspace{5mm}  0.3 &\hspace{5mm} 9.94 & 0.576 & 0.027 &\hspace{5mm} 3.392 \\
  & \hspace{5mm}  0.9 &\hspace{5mm}  &\hspace{5mm} 10.36 & 0.472 & 0.025 &\hspace{5mm} 3.249 \\
  & \hspace{5mm}  1.0 &\hspace{5mm}  &\hspace{5mm} 10.78 & 0.408 & 0.024 & \hspace{5mm} 3.121 \\

  \hline
  \end{tabular}
\end{table*}

In general the angular velocity of a test particles could be obtained through the geodesic equations (\ref{pr})-(\ref{pt}) and expressed as a function of the metric components (\ref{mc}), they are
\begin{equation}\label{eq31}
 \Omega_{\pm}=\frac{\dot{t}}{\dot{\phi}}=\frac{-g_{t\phi,r}\pm\sqrt{(g_{t\phi,r})^{2}-g_{tt,r}g_{\phi\phi,r}}}{g_{\phi\phi,r}}\,,
\end{equation}
the upper and lower signs indicate co-rotating and counter-rotating orbits, respectively. Applying equations (\ref{eq24}) and (\ref{eq25}), the specific energy $E$ and specific angular momentum $L$ of the test particles moving in circular orbits can be calculated as follws
\begin{equation}\label{eq32}
 E=-\frac{g_{tt}+g_{t\phi}\Omega}{\sqrt{-g_{tt}-2g_{t\phi}\Omega-g_{\phi\phi}\Omega^{2}}},
\end{equation}
\begin{equation}\label{eq33}
 L=\frac{g_{t\phi}+g_{\phi\phi}\Omega}{\sqrt{-g_{tt}-2g_{t\phi}\Omega-g_{\phi\phi}\Omega^{2}}}.
\end{equation}
Using equations (\ref{mc}), we can calculate the angular velocity, specific energy, specific angular momentum and effective potential of particles in circular orbits around an electrically charged rotating EEH black hole. Due to complexity and lengthy of these equations, it is not suitable to write them, instead one can analyze their behaviour through numerical study and their plots.

\begin{figure}%
    \centering
    \subfloat[\centering A=0.99,Q=0.3]{{\includegraphics[width=8cm]{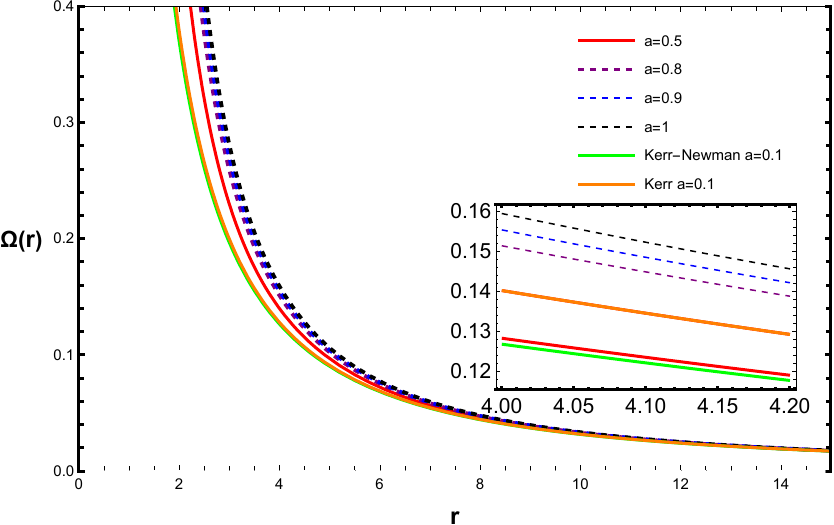} }}%
    \qquad
    \subfloat[\centering A=0.99,a=0.3]{{\includegraphics[width=8cm]{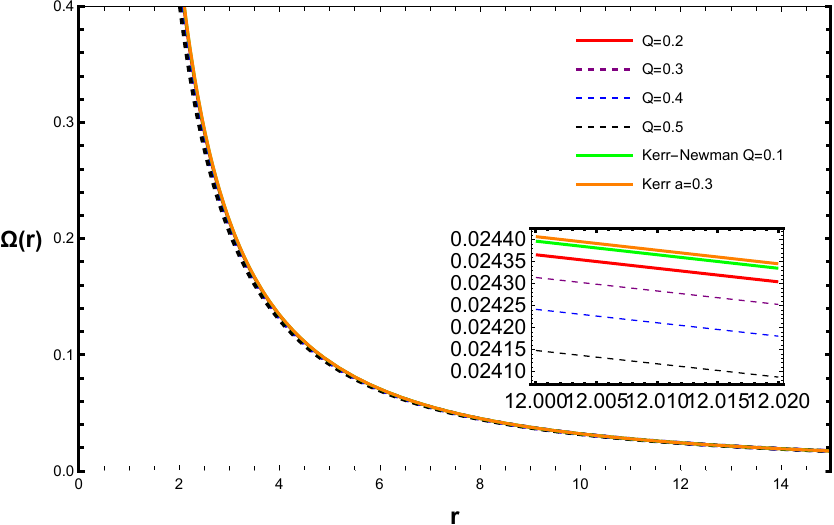} }}%
    \caption{The angular velocity of the electrically charged rotating EEH black hole as a function of the radial coordinate $r$ for different values of the spin parameter and $Q$. }%
    \label{fig2}%
\end{figure}

\begin{figure}%
    \centering
    \subfloat[\centering A=0.99,Q=0.3]{{\includegraphics[width=8cm]{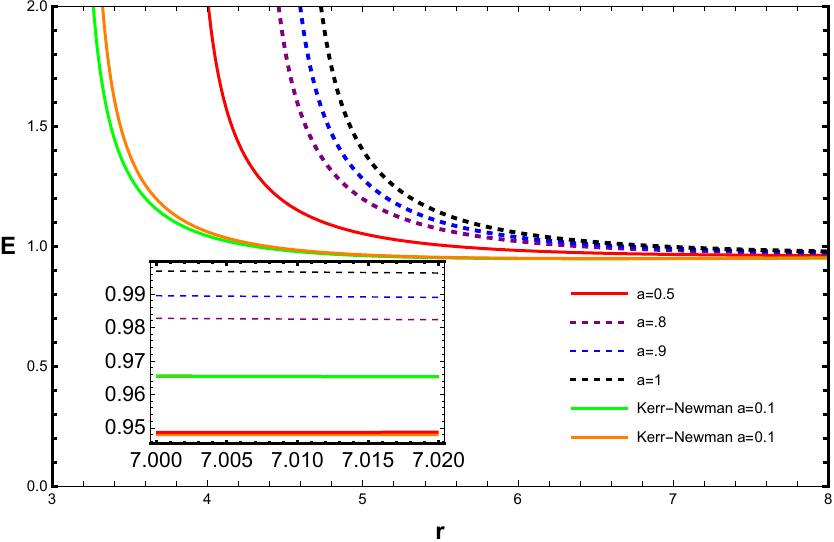} }}%
    \qquad
    \subfloat[\centering A=0.99,a=0.3]{{\includegraphics[width=8cm]{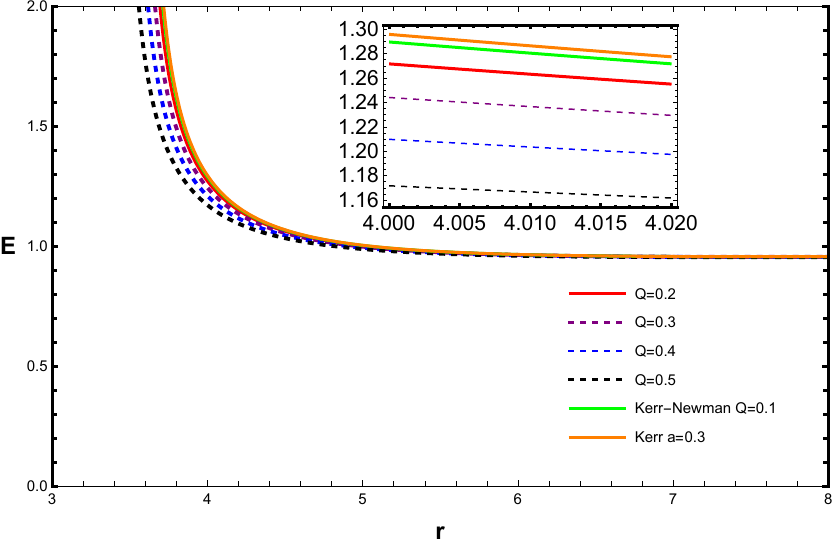} }}%
    \caption{The specific energy of the electrically charged rotating EEH black hole as a function of the radial coordinate $r$ for different values of the spin parameter and $Q$.}%
    \label{fig3}%
\end{figure}

\begin{figure}%
    \centering
    \subfloat[\centering A=0.99,Q=0.3]{{\includegraphics[width=8cm]{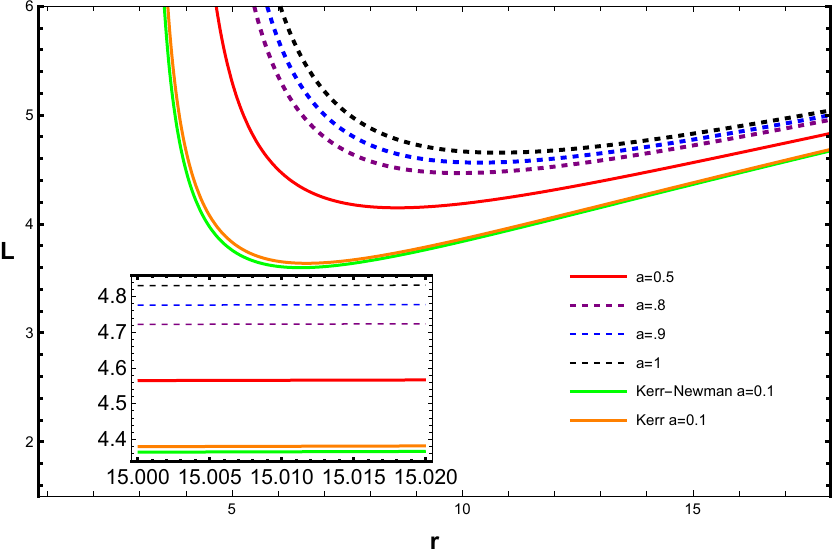} }}%
    \qquad
    \subfloat[\centering A=0.99,a=0.3]{{\includegraphics[width=8cm]{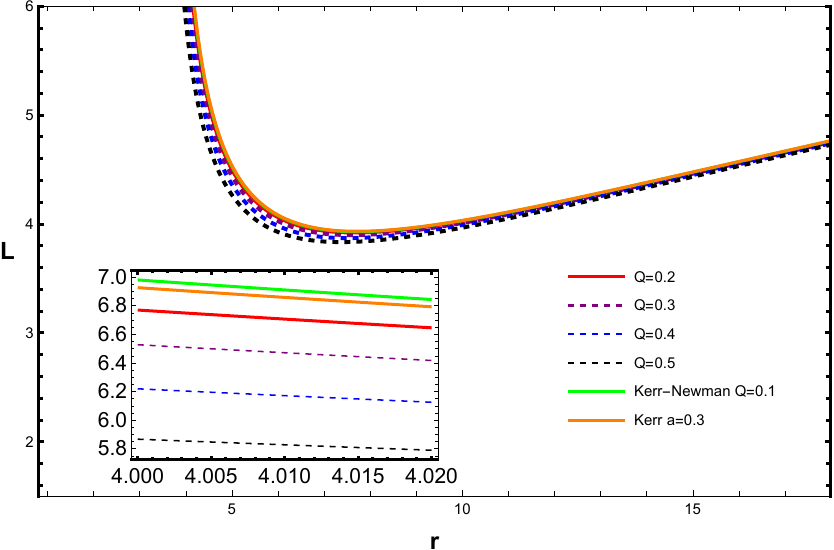} }}%
    \caption{The angular momentum of the electrically charged rotating EEH black hole as a function of the radial coordinate $r$ for different values of the spin parameter and $Q$.}%
    \label{fig4}%
\end{figure}

Figures \ref{fig2}, \ref{fig3} and \ref{fig4} show the behaviour of the specific energy, specific angular
momentum and angular velocity of the electrically charged rotating EEH black hole for different values of the spin parameter and electric charge. They also contain a comparison with the results obtained for Kerr and Kerr-Newman black holes in GR. We see that by fixing the value of $Q$, increasing the spin parameter causes the above quantities to increase in comparison to Kerr and Kerr-Newman black holes. While for a fixed $\textmd{a}$, increasing the charge $Q$ causes them to decrease compared to those of Kerr and Kerr-Newman black holes.

Table \ref{t1} shows that for a fixed value of the electric charge, the effect of spin parameter is to increase the ISCO radius of the electrically charged rotating EEH black hole, and this effect is different from that in a standard Kerr black hole due to the additional electromagnetic corrections introduced by the Euler-Heisenberg theory. These corrections can affect the structure of the black hole's event horizon and the properties of orbits near the black hole horizon. Also, the ISCO radius of the electrically charged rotating EEH black hole is larger than the ISCO radius in a Kerr black hole. In particular, for a rotating EEH black hole, the ISCO is often located farther away from the event horizon compared to a Kerr black hole. While, by fixing the spin parameter the ISCO radius of the electrically charged rotating EEH black hole decreases, due to the increase in the electric charge.

\section{ Radiative properties of thin accretion disk in EEH Theory}\label{se4}

In this analysis, we adopt the steady-state Novikov-Thorne framework to investigate the structure and dynamics of a geometrically thin accretion disk surrounding an electrically charged rotating EEH black hole. This model describes a disk whose vertical scale height ($h$) is significantly smaller than its radial extent ($r$), satisfying the condition $h \ll r$. The disk is confined to the equatorial plane of the central object, and its inner edge coincides with the radius of the innermost stable circular orbit (ISCO). Between this inner boundary, denoted as $r_{\rm isco}$, and the outer radius, $r_{\rm out}$, the disk matter undergoes nearly circular Keplerian motion. Moreover, the mass accretion rate, $\dot{M}$, is considered to be steady and uniform throughout the disk over time \cite{bombi2017}.

\begin{figure}%
    \centering
    \subfloat[\centering A=0.99,Q=0.3]{{\includegraphics[width=8cm]{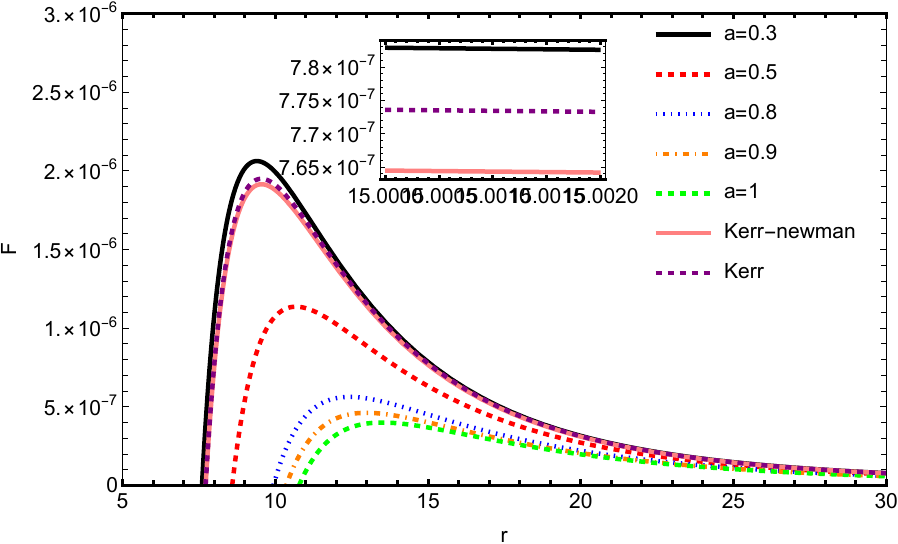} }}%
    \qquad
    \subfloat[\centering A=0.99,a=0.3]{{\includegraphics[width=8cm]{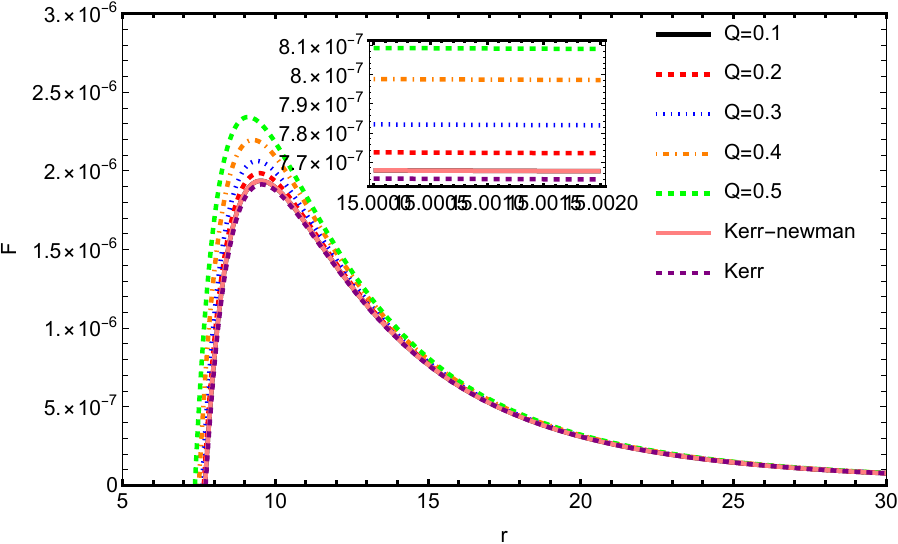} }}%
    \caption{ The energy flux $F$ versus radial coordinate $r$ of an accretion disk around the electrically charged rotating EEH black hole with the mass accretion rate $\dot{M}=2\times10^{-6}M_{\odot}\rm yr^{-1}$ for different values of the spin parameter and $Q$. }%
    \label{fig5}%
\end{figure}

The radiant energy flux emitted by the accretion disk can be derived using the conservation laws of energy $(\nabla_\mu T^{t\mu} = 0)$, angular momentum $(\nabla_\mu T^{\phi\mu} = 0)$, and rest mass $(\nabla_\mu(\rho u^\mu) = 0)$ for the disk particles, as outlined in \citep{novikov1973black,Page:1974he}
\begin{equation}
F(r)=-\frac{\dot{M}\Omega_{,r}}{4\pi\sqrt{-g}(E-\Omega L)^2}\int_{r_{\rm isco}}^r (E-\Omega L)L_{,r}dr.
\end{equation}
This equation expresses the radial flux of energy emitted from the accretion disk surface around a rotating compact object, according to the Novikov-Thorne thin disk model. $F(r)$ is a flux that depends on the radial derivative of $\Omega$, involving an integral from ${r_{\rm isco}}$ to  $r$ that describes how angular momentum and energy are redistributed within the disk. The factor $(E-\Omega L)^2$ accounts for the redshifting and frame dragging imposed upon the emitted radiation by the strong gravitational field.
$\dot M $ is the mass accretion rate,
$g$ is the metric determinant of three dimensional subspace $(t,r,\phi)$
\begin{equation}
g = g_{rr}(g_{tt}g_{\phi\phi}-g_{t\phi}^2).
\end{equation}
Eqs.~(\ref{eq31}), (\ref{eq32}), and (\ref{eq33}) provide values for $\Omega$, $E$, and $L$, respectively. $\Omega_{, r}$ and $L_{, r}$ are the radial derivatives of angular velocity and angular momentum of a test particle. In Fig ~{\ref{fig5}}, we demonstrate the dependency of radiative flux for various black hole spins $\textmd{a}$ and electric charge $Q$. The energy flux distribution shows a unique pattern of initial rise, peak attainment, and subsequent fall. Furthermore, we see that, retaining constant the charge $Q$, as well as the distance $r$, the radiation flow drops as the spin parameter increases. However, leaving the spin parameter unchanged, increasing the electric charge results in a gradual increase in radiation flux $F(r)$. Also, the figure provides a comparison between the results obtained for Kerr and Kerr-Newman black holes in GR. Moreover, the inner radii at which the radiative flux approaches zero correspond to
$r_{\rm isco}$. It can be observed that these radii shift to smaller values with increasing the electric charge and decreasing the spin parameter of the electrically charged rotating EEH black hole. In the distant region, all the figures of energy flux appear to converge, remaining unaffected by the black hole spin and the electric charge. Additionally, by fixing $Q$, it becomes evident that the electrically charged rotating EEH black hole exhibits a smaller energy flux compared to the Kerr and Kerr-Newman black holes. However, increasing the electric charge enhances the energy flux of the electrically charged rotating EEH black hole, shifting its peaks toward smaller radii. Consequently, a significant portion of the energy flux is emitted from the inner regions of the disk.

\begin{figure}%
    \centering
    \subfloat[\centering A=0.99,Q=0.3]{{\includegraphics[width=8cm]{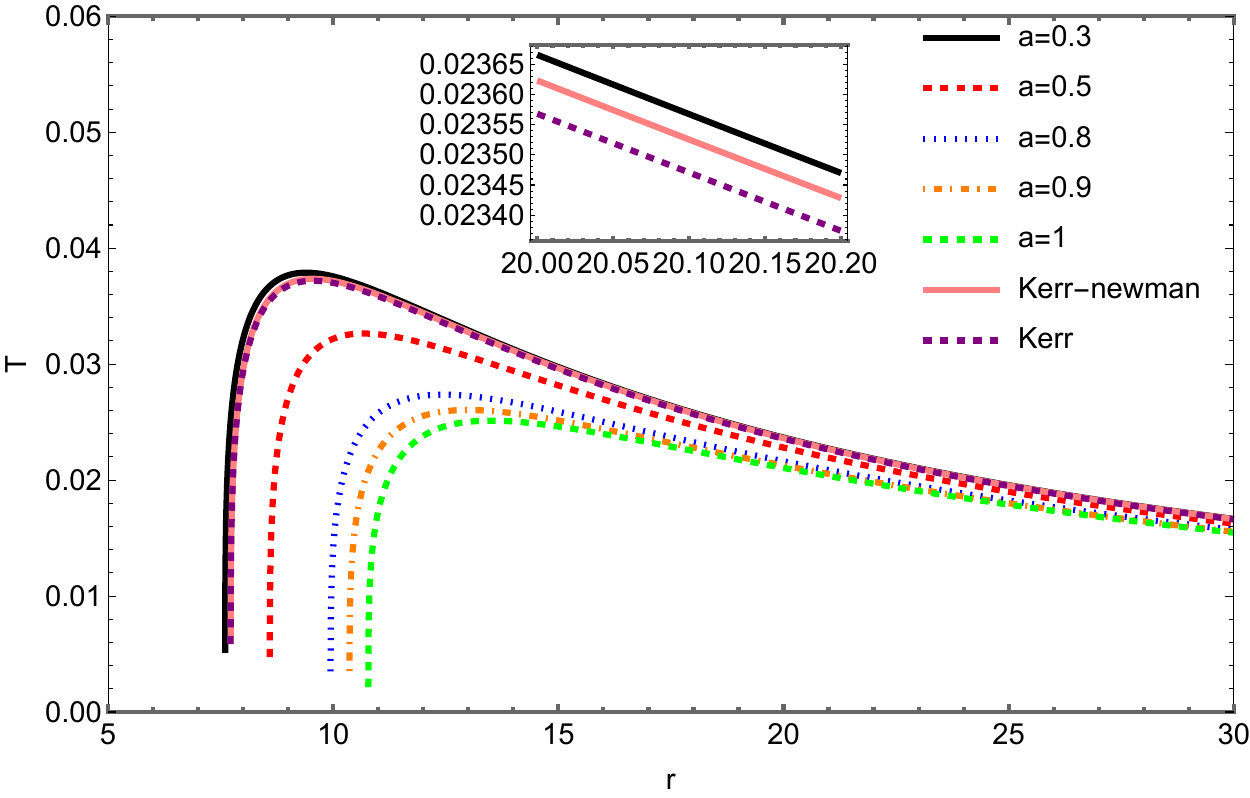} }}%
    \qquad
    \subfloat[\centering A=0.99,a=0.3]{{\includegraphics[width=8cm]{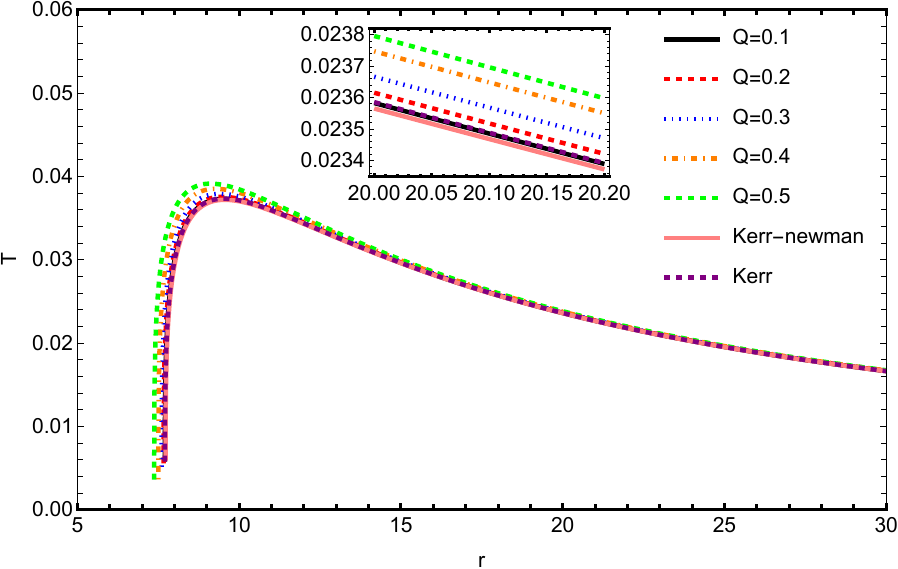} }}%
    \caption{The radiation temperature $T$ versus radial coordinate $r$ of an accretion disk around the electrically charged rotating EEH black hole with the mass accretion rate $\dot{M}=2\times10^{-6}M_{\odot}\rm yr^{-1}$ for different values of the spin parameter and $Q$.}%
    \label{fig6}%
\end{figure}

Within the Novikov-Thorne framework, the accreted material reaches thermodynamic equilibrium, resulting in the disk's radiation behaving like that of a perfect black body. The disk's radiation temperature, $T(r)$, is related to the energy flux, $F(r)$, through the Stefan-Boltzmann law, given by $F(r)=\sigma_{SB}T^{4}(r)$, where $\sigma_{SB}$ is the Stefan-Boltzmann constant. This relationship implies that the variation of $T(r)$ with $r$ mirrors the dependence of $F(r)$ on $r$. The radiation temperature $T(r)$ around the electrically charged rotating EEH black hole is depicted in Fig. \ref{fig6}. Additionally, we systematically vary one parameter of the charged rotating EEH black hole while keeping the other fixed to analyze its effect on $T(r)$. From the left plot, it is clear that as the spin parameter increases, the temperature $T(r)$ decreases. Similarly, the right plot, corresponding to $A = 0.99$ and $\textmd{a} = 0.3$, shows that an increase in the electric charge results in an increase in $T(r)$. The overall temperature profile exhibits a behavior where it initially increases and subsequently decreases. At larger radii, the temperature profiles converge, indicating the diminishing influence of both the black hole spin $a$ and the electric charge $Q$ in the distant region. As illustrated in Fig \ref{fig6}, the accretion disk surrounding the electrically charged rotating EEH black hole exhibits a lower temperature compared to the disks of Kerr and Kerr-Newman black holes for a fixed electric charge $Q$. However, an increase in the electric charge $Q$ results in an elevated temperature of the accretion disk of the electrically charged rotating EEH black hole, highlighting the influence of the charge parameter on the thermal structure of the disk.\\

It is important to note that the radiative flux $F(r)$ is not directly observable, as it is defined in the rest frame of the accretion disk. In contrast, from an observational standpoint, the differential luminosity $L_{\infty}$ is a more relevant and practical quantity, as the radiative flux cannot be directly measured. As outlined in Refs.~\cite{Bosh:2023,Josh:2013}, the differential luminosity is given by
\begin{equation}
 \label{eq:difflum}
\frac{dL_{\infty}}{d\ln{r}}=4\pi r \sqrt{-g}E F(r),
\end{equation}

\begin{figure}%
    \centering
    \subfloat[\centering A=0.99,a=0.3]{{\includegraphics[width=8cm]{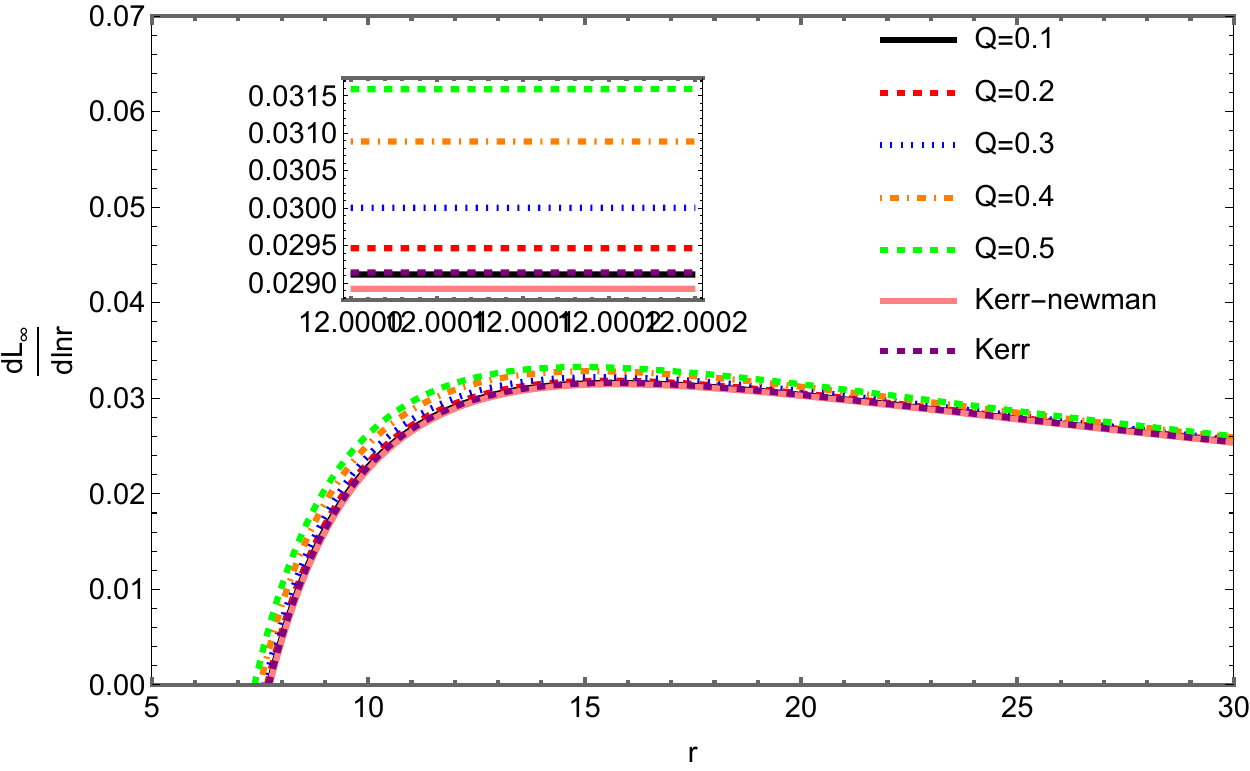} }}%
    \qquad
    \subfloat[\centering A=0.99,Q=0.3]{{\includegraphics[width=8cm]{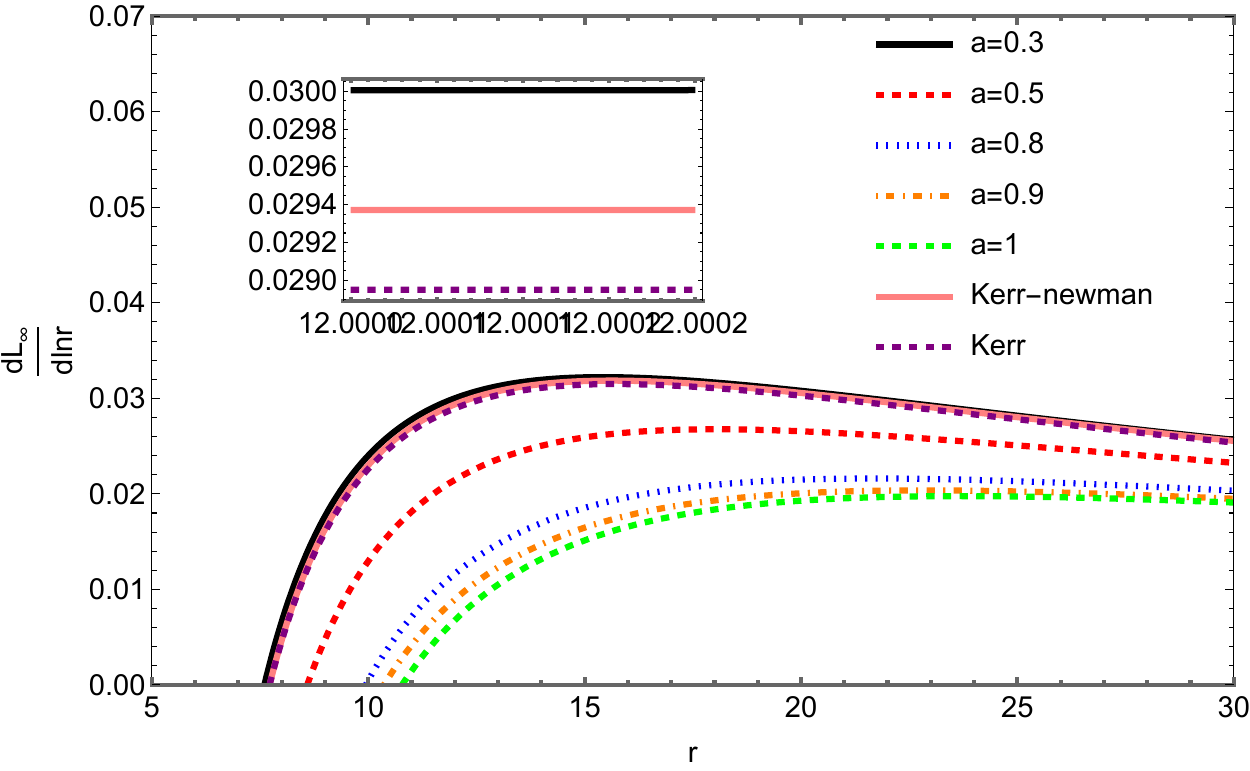} }}%
    \caption{The differential luminosity $\frac{dL_{\infty}}{d\ln{r}}$ versus radial coordinate $r$ of an accretion disk around the electrically charged rotating EEH black hole with the mass accretion rate $\dot{M}=2\times10^{-6}M_{\odot}\rm yr^{-1}$ for different values of the spin parameter and $Q$. }%
    \label{fig7}%
\end{figure}

The differential luminosity clearly describes the radiation emitted by the accretion disk at a given distance. From the differential luminosity, it is possible to derive the spectrum and frequencies, which represent directly measurable quantities. In Fig.~\ref{fig7}, we present the differential luminosity as a function of the radial coordinate for the electrically charged rotating EEH black hole, where we fix $Q=0.3$ and vary the spin parameter a (left panel), and fix $\textmd{a}=0.3$ while varying $Q$ (right panel). We observe a behavior of the differential luminosity that closely mirrors the radiative flux shown in Fig.~\ref{fig5}. This similarity arises from the fact that both quantities are related through Eq.~(\ref{eq:difflum}), meaning that the trends observed in the flux are directly reflected in the differential luminosity.\\

Finally, let us define the accretion efficiency $\epsilon$, which is defined as the ratio of the energy radiated away during the accretion process to the rest-mass energy of the infalling material. This efficiency is fundamental in determining both the luminosity and spectral properties of the accretion disk. It is often measured in terms of the Eddington luminosity, which represents the maximum luminosity a black hole can reach when the outward radiation pressure balances the inward gravitational force. Understanding the radiative efficiency is crucial for comprehending the accretion dynamics of black holes, as well as their effects on the surrounding environment and their observational signatures across different wavelengths. Assuming that all emitted photons can escape to infinity, the efficiency $\epsilon$ is determined by the specific energy of a particle at the marginally stable orbit $r_{\rm isco}$, as expressed in \cite{Page:1974he}
\begin{equation}
\epsilon = 1 - E_{\rm isco}.
\end{equation}

In Table~\ref{t1}, we present the numerical results for maximal values of the energy flux, temperature distribution and the radiative efficiency of the electrically charged rotating EEH black hole, calculated for various values of the spin parameter and charge.

\section{Conclusions}\label{se5}

Accretion disks around supermassive black holes serve as the primary source of gravitational information in strong-field regimes and provide crucial insights into the surrounding space-time geometry. Previous studies \citep{Rehman:2023hro,Abbas:2023nra} have investigated the accretion disk around the Euler-Heisenberg Anti-de
Sitter static black hole, as well as matter accretion onto the static magnetically charged Euler-Heisenberg black hole with
scalar hair. Given that astrophysical black holes are expected to acquire rotation through accretion processes, we extend this analysis to the case of electrically charged, rotating black holes within the framework of Euler-Heisenberg (EH) theory. By incorporating QED corrections via the Euler-Heisenberg Lagrangian, our study demonstrates significant deviations from the standard predictions of Kerr and Kerr-Newman black holes in general relativity, highlighting the impact of QED effects on the dynamics of accretion disks in strong gravity environments.

In this study, we employed the Novikov-Thorne model to examine the properties of thin relativistic accretion disk surrounding the electrically charged rotating EEH black hole. As a first step, we derived key physical quantities of interest, including the effective potential $V_{eff}$, specific angular momentum $L$, specific energy $E$, angular velocity $\Omega$ and the ISCO radius for test particles moving in circular orbits around the black hole. Since the ISCO equation does not admit an analytical solution, we determined the ISCO radius numerically, considering the effects of both the spin parameter and the electric charge. The QED corrections inherent in the EEH framework cause the ISCO radius to increase  compared to those of Kerr and Kerr-Newman black holes. For a given electric charge, a higher spin parameter results in a larger ISCO radius than in standard Kerr black holes due to the extra electromagnetic corrections from Euler-Heisenberg theory. On the other hand, if the spin parameter remains fixed, increasing the electric charge causes the ISCO radius to decrease. This highlights the profound influence of non-linear electrodynamic effects on particle dynamics and accretion processes near black holes.

To analyze the radiative properties of thin accretion disks around electrically charged rotating EEH black holes, we numerically computed the energy flux $F(r)$, temperature $T(r)$ and differential luminosity $L_{\infty}$ and analyzed their profiles. The increase in the ISCO radius suggests that the gravitational field intensifies with a higher spin parameter due to QED effects. As a result, the inner edge of the accretion disk, which is defined by the ISCO radius in the Novikov-Thorne model, shifts farther from the event horizon. Consequently, in comparison to Kerr and Kerr-Newman black holes, the energy flux, temperature, and differential luminosity of the disk are reduced, which is consistent with the results presented in Table~\ref{t1}. However, for a fixed spin parameter, an increase in the electric charge enhances the radiative flux and temperature, leading to a hotter and more efficient accretion disk. Additionally, the radiative efficiency of accretion disks around the charged rotating EEH black hole was found to be lower compared to Kerr solutions, which could have observational implications.

In conclusion, this work underscores the importance of considering non-linear electrodynamics and alternative gravity theories to enhance our understanding of the extreme environments around black holes. The deviations observed in ISCO radii, radiative efficiency, and temperature profiles offer valuable insights into the impact of QED corrections on black hole physics. These differences could serve as potential observational signatures for detecting and distinguishing EEH black holes in astrophysical scenarios.\\

\begin{acknowledgments}

The Work of K. Nozari and S. Saghafi is supported financially by the INSF of Iran under the grant number $4038520$.

\end{acknowledgments}

\end{document}